\documentstyle[12pt,aasms4,overcite,flushrt]{article}

\renewcommand{\thebibliography}[1]{\clearpage\subsection*{REFERENCES}\list
 {\arabic{enumi}.}{\settowidth\labelwidth{[#1]}\leftmargin\labelwidth
 \advance\leftmargin\labelsep
 \usecounter{enumi}}
 \def\newblock{\hskip .11em plus .33em minus .07em}
 \sloppy\clubpenalty4000\widowpenalty4000
 \sfcode`\.=1000\relax}

\def\aasup#1{{\it Astr. Astrophys. Suppl. Ser.\/} {\bf #1}}
\def\aj#1{{\it Astr. J.\/} {\bf #1}}
\def\apj#1{{\it Astrophys. J.\/} {\bf #1}}
\def\apjsup#1{{\it Astrophys. J. Suppl. Ser.\/} {\bf #1}}

\def\mn#1{{\it Mon. Not. R. astr. Soc.\/} {\bf #1}}

\def\nat#1{{\it Nature\/} {\bf #1}}

\def\bmi{{$(B-I)$}}
\def\etal{{\it et~al.\/}}
\def\feii{{\rm Fe}\thinspace{\sc{ii}}}
\def\ha{\ifmmode {{\rm H}\alpha}
        \else {H$\alpha$}\fi}
\def\hb{\ifmmode {{\rm H}\beta}
        \else {H$\beta$}\fi}
\def\hnought{\ifmmode H_0
    \else $H_0$\fi}
\def\imk{{$(I-K)$}}
\def\kms{~{\rm km\ s}^{-1}}
\def\lstar{{$L_{\ast}$}}
\def\mdot{\rm M_{\odot}\>yr^{-1}}
\def\mgii{{\rm Mg}\thinspace{\sc{ii}}}
\def\mnii{{\rm Mn}\thinspace{\sc{ii}}}
\def\msun{\ifmmode {\rm M_{\odot}}
       \else ${\rm M_{\odot}}$\fi}
\def\oii{[O\thinspace{\sc{ii}}]}
\def\oiii{[O\thinspace{\sc{iii}}]}
\def\p.{^{\prime\prime}\kern-2.1mm .\kern+.6mm}
\def\qnought{\ifmmode q_0
    \else $q_0$\fi}
\def\ten#1{\ifmmode 10^{#1}
    \else $10^{#1}$\fi}
\begin{document}

\title{Detection of massive forming galaxies\\ at redshifts greater than one}
\author{Lennox L. Cowie, Esther M. Hu \& Antoinette Songaila}
\affil{Institute for Astronomy, University of Hawaii, 2680 Woodlawn Dr.,
  Honolulu, HI 96822\\}

\vskip 1in

\centerline{Accepted for publication in {\it Nature}}

\newpage
{\bf The complex problem of when and how galaxies formed has not until
recently been susceptible of direct attack.  It has been known for some
time that the excessive number of blue galaxies counted at faint
magnitudes\cite{kron,koo,tyson,lcg,metcalfe95} implies that a
considerable fraction of the massive star formation in the universe
occurred at \mbox{\boldmath $z<$ 3}\cite{cowie88,songaila90}, but,
surprisingly, spectroscopic studies\cite{broad88,colless90,csh,glaze95}
of galaxies down to a \mbox{\boldmath $B$}\ (blue) magnitude of 24 found
little sign of the expected high-{\boldmath $z$}\ progenitors of current
massive galaxies, but rather, in large part, small blue galaxies at
modest redshifts (\mbox{\boldmath $z\sim$ 0.3}).  This unexpected
population has diverted attention from the possibility that early massive
star-forming galaxies might also be found in the faint blue excess.  From
our first spectroscopic observations deep enough to encompass a large
population of \mbox{\boldmath $z>$ 1} field galaxies, we can now show
directly that in fact these forming galaxies are present in substantial
numbers at \mbox{\boldmath $B\sim$ 24}, and that the era from redshifts 1
to 2 was clearly a major period of galaxy formation.  These
\mbox{\boldmath $z>$ 1}\ galaxies have very unusual morphologies.}

The recent availability of the highly sensitive multi-object Low
Resolution Imaging Spectrograph (LRIS) on the Keck 10 m telescope on Mauna
Kea, Hawaii has allowed us to extend and deepen quite considerably our
spectroscopic coverage of the faint sky, and we are currently carrying out
a redshift survey of faint galaxies using this instrument.  The sample
will consist of all objects satisfying the conditions $K<20$\ or $I<22.5$
or $B<24.5$ in four $6'\times2'$ fields; so far, the most complete are
those surrounding the Hawaii deep field SSA13\cite{lcg,paper1} in which
147 of the 174 objects have been observed and 124 of these securely
identified, and around SSA22, with 186 of the 193 objects observed and 157
identified.  In all, ninety-one galaxies lying beyond $z=0.7$ and forty
beyond $z=1$ have been identified in these two fields, with the highest
redshift being $z=1.69$.  We show (Fig.~1) representative spectra of four
galaxies at $z>1$.  Only three of the $z>0.7$ sample are galaxies with
active nuclei (AGN).  Most of the high redshift identifications are based
on the presence of a strong \oii\ 3727~\AA\ emission line (a nebular
emission line of singly ionized oxygen) and weak \mgii\ 2800~\AA\
absorption\cite{glaze95}.  The general form of the spectra is
illustrated in more detail in Fig.~2, which is the average of all the
$z>0.7$ spectra in SSA13, excluding only the AGN.  The composite spectrum
closely resembles that of local blue galaxies\cite{rosa,lamb,york} with
very strong \oii\ 3727, \hb\ and \oiii\ 5007, 4959 emission lines in the
optical, and UV absorption lines of \feii, \mnii\ and \mgii\ that confirm
the \oii\ redshift identification.  These absorption lines in the
composite spectrum are redshifted by an average of $300 \kms$ from the
emission lines, and may arise from gas infalling into the galaxy.

The \oii\ 3727 emission lines in the high-redshift objects are remarkably
strong (rest equivalent width of 60 \AA\ in the composite spectrum) given
that these are luminous galaxies with absolute rest $B$ magnitudes of about
$-21$ for \hnought=$50\kms$ Mpc$^{-1}$ and \qnought=0.5.  The \oii\
luminosities ($L_{\rm [O\thinspace{II}]}$) of the individual galaxies in
the two fields are shown in Fig.~3(a), and we have also shown as crosses
the \oii\ luminosities for the Hawaii $K\leq19$ sample.\cite{paper3} At low
redshift ($z<0.7$) there are no galaxies in either sample with $L_{\rm
[O\thinspace{II}]} > \ten{42}$ ergs s$^{-1}$ while at $z>0.7$ there are
27 objects identified in these two fields.  It is important to note
that it is much easier to recognize and measure redshifts for objects with
strong \oii\, and there may be $z>1$ objects with lower \oii\ luminosities
hidden in the unidentified objects.  There may also be additional objects
at these redshifts lying beyond the magnitude limits of the present sample
or among the remaining unobserved objects.  This means the sample is not
complete, but the key point is that even in this incomplete sample there
is already a very large surface density of luminous star-forming galaxies.

To quantify this we must translate the \oii\ luminosities into galaxy
formation rates. This is discussed by Gallagher \etal\cite{gallagher} and
by Kennicutt\cite{kenn}, who point out that the \oii\ line, though a more
indirect calibrator than the unobserved \ha, does provide a useful estimate
of the star formation rate, expressed in solar masses (\msun) per year as:
\begin{equation}
	\dot{M}\,({\mdot}) = \ten{-41}\ \alpha\ L_{\rm [O\thinspace{II}]}\
	(\rm ergs\ s^{-1}).
\end{equation}
We have introduced the parameter $\alpha$ to reflect the significant
uncertainties in \oii/\ha\ ratios, internal extinction, and the assumed
stellar initial mass function (IMF).  Gallagher \etal\ find $\alpha=1$
based on a sample of nearby blue galaxies while Kennicutt finds $\alpha=5$
for a wider sample of types.  This range of values is a good measure of the
uncertainty but the lower Gallagher \etal\ value may be more representative
of the present sample, which is extremely blue.

For \hnought=$50\kms$ Mpc$^{-1}$ and \qnought=0.5, as assumed in Fig.~3,
the age of the universe at $z=1$ is $6\times \ten{9}h^{-1}$ yrs ($h =
\hnought /50\kms {\rm Mpc}^{-1}$) and formation of a `normal' galaxy with
$6\times\ten{10}h^{-1}\ \msun$ of stars (a so-called \lstar\ galaxy)
would require a continuous star formation rate of 10 $\mdot$,
corresponding to $L_{\rm [O\thinspace{II}]} = \ten{42}$ ergs s$^{-1}$ for
$\alpha=1$\ (the dashed line in Fig.~3(a)).  We note again that such
galaxies are not seen at $z<0.7$ in the field samples (Fig.~2) nor in
either of the local\cite{gallagher,kenn} samples, but are extremely
common at $z>0.7$.  However, star formation need not proceed at a uniform
rate in any individual galaxy, so the present observations might be
picking out only the small subsample of starbursting galaxies at these
redshifts.  The best way to resolve this is to consider the volume
production rate of \oii\ photons by the ensemble of galaxies and compare
this to the present mass density of stars in galaxies.  This ensemble
production rate of \oii\ or \ha\ line photons directly measures the rate
of massive star production and so can be linked immediately to the
present day density of metals{\cite{cowie88,songaila90}} in the universe
or, slightly more indirectly and subject to uncertainties in the IMF, to
the present-day mass density of stars in the universe.

We first define the volume production rate of \oii\ photons as
\begin{equation}
	{\cal L}_{\rm [O\thinspace{II}]}(z) = \sum_{\Delta z}
 { {\displaystyle{L_{\rm [O\thinspace{II}]}}}\over{\displaystyle{V_{\Delta
z}}}}
\end{equation}
where the sum is over all galaxies lying in the redshift interval $\Delta z$
surrounding $z$, and $V_{\Delta z}$ is the comoving volume corresponding to
the observed area.  {\it De facto\/} we are restricted in the summation to
observed, identified galaxies and our measured values are lower bounds, with
the correction being larger at high $z$.  ${\cal L}_{\rm
[O\thinspace{II}]}(z)$, which can be converted directly to a stellar mass
density formation rate using equation (1), is shown in Fig.~3(b) for redshift
intervals from $z=0.25$ to 1.5.  (It scales as $h$.) We have distinguished
entries for SSA13 (boxes) and SSA22 (diamonds) to give some feeling for the
uncertainty, but evidently the total rate of (massive) star formation was much
higher in the recent past: even without any further correction for missing
objects the star formation rate at $z=1.125$ was four times higher than that
at $z=0.375$.

Turning to absolute values, we can compare the measured star formation rates
with that required to form the presently observed star density.  For
reference purposes we assume that the current stellar mass density is
$3\times\ten{8}h^2\ \msun\ {\rm Mpc}^{-3}$\ and that the available time is
$17\ h^{-1}\ (1+z)^{-3/2}$ Gyr.  To form the current stars at some period
then requires a star formation rate,
\begin{equation}
	\dot\mu_{ref} = 1.7 \times \ten{-2}\ {(1+z)^{3/2}}\ h^3
	\ {\rm M_{\odot}\>Mpc^{-3}\>yr^{-1}}.
\end{equation}
This reference rate is shown as the dashed and dotted lines in Fig.~3(b),
which are computed for the values $\alpha=1$ and $\alpha=5$ respectively
in equation 1.  At $z>1$, even in this quite incomplete sample, we are
already seeing between 4\% and 20\% of the total galaxy formation.

Most of the luminous \oii\ galaxies and most of the identified $z>1$ galaxies
are drawn from objects with blue optical colors [\bmi$\leq1.7$] but
relatively red infrared colors\cite{paper1} [\imk$\geq1.8$] (Fig.~4).  Eleven
of the sixteen objects in SSA13 with $L_{\rm [O\thinspace{II}]}\geq\ten{42}$
ergs s$^{-1}$ and thirteen of the twenty objects with $z>1$ lie in this portion
of the color-color plane; conversely nearly all the identified objects in
this region are predominantly \oii-luminous high-$z$ galaxies.  Of the 18
objects identified in these regions two are AGN and two are low-$z$ galaxies
but all those remaining have $L_{\rm [O\thinspace{II}]}>4\times\ten{41}$ ergs
s$^{-1}$ and lie at $z>0.7$.  Inspection of the color-color plane suggests
that nearly all the low-$z$ galaxies in the sample have been identified and
the remaining unidentified or unobserved objects are luminous high-$z$ star
formers.  Completion of the sample should therefore increase the $z>1$ galaxy
formation rate substantially, bringing the measured mass rate much closer to
the reference value.

Finally we have used deep $I$-band images, obtained with the wide field
camera (WFPC2) on the Hubble Space Telescope (HST), of the central regions
of the SSA13 and SSA22 fields\cite{chain} to investigate the morphologies
of the high-$z$ starbursters.  The images of all nine known objects in the
HST fields with $z > 1$\ and $L_{\rm [O\thinspace{II}]}\geq\ten{42}$ ergs
s$^{-1}$ are shown in Fig.~5.  Unlike lower redshift objects with high
\oii\ equivalent widths\cite{coll94} the objects have strikingly unusual
morphologies, often consisting of chains\cite{chain} or structures of
compact blobs, suggesting that they are generally not dominated by
uniformly distributed star formation.  Recent studies have
shown\cite{chain,kgb} that the excess galaxy counts at faint blue
magnitudes are dominated by these anomalous galaxies and that at faint
infrared magnitudes ($K>20$) these also become the single largest
population\cite{chain}.  Combined with the present results we can now
recognize that forming galaxies enter the galaxy counts in substantial
numbers at faint ($B>24$, $K>20$) magnitudes.

\smallskip
\acknowledgments
We would like to thank K. Glazebrook for a critical reading of a first draft
of this manuscript and M. Colless, J. Cohen, J. Gallagher, O. LeF\`evre and R.
Williams for comments on the final version.  We are grateful to T. Bida, P.
Gillingham, J. Aycock, T. Chelminiak and W. Wack for their extensive help in
obtaining the observations, which would not have been possible without J.
Cohen and B. Oke's LRIS spectrograph.  The authors were visiting astronomers
at the W. M. Keck Observatory, jointly operated by the California Institute
of Technology and the University of California.  This work was partly based
on observations with the NASA/ESA {\it Hubble Space Telescope\/} obtained at
the Space Telescope Science Institute, which is operated by AURA, Inc., under
NASA contract. The research was supported at the University of Hawaii by the
State of Hawaii and by NASA.

\newpage

\centerline{\bf FIGURES}
\begin{figure}[h]
\caption{Sample spectra of $z> 1$\  galaxies observed with LRIS on the
Keck 10 m telescope.  Each object was observed with a $1\p.4$ wide slit
using the 300$\ell$/mm grating giving a resolution of 17 \AA\ and a
wavelength coverage of 5000~\AA.  Exposure times range from one to four
hours.  Each object is shown as $f_{\nu}$ versus observed wavelength with
the shaded regions showing the positions of strong night sky lines and
atmospheric absorption bands.  The tick marks show the position of \mgii\
2800 \AA\ absorption and \oii\ 3727 \AA\ emission at the redshift shown
in the lower left corner.  The apparent Kron-Cousins $I$ magnitude is
shown at the lower right.}
\end{figure}

\begin{figure}[h]
\caption{The average rest wavelength spectrum of objects with $z>0.7$\ in
the SSA13 field (excluding AGN).
This composite spectrum was formed by normalizing all the spectra to their
median value and then averaging all the spectra that covered a particular
wavelength region.  The optical portion of the spectrum is characterized by
strong emission lines but only weak absorption lines are seen in the UV.}
\end{figure}

\begin{figure}[h]
\caption{(a)\ The \oii\ luminosities of galaxies in the SSA13 field (boxes)
and SSA22 field (diamonds), computed by normalizing each spectrum to the
flux in the measured broad-band magnitude nearest to the redshifted \oii\
line to determine the continuum flux per unit wavelength, $f_{\lambda}$ and
combining this with the luminosity distance at redshift $z$.  The Songaila
\etal\ (1995) sample are also shown as crosses.  The dashed line roughly
divides
rapidly forming galaxies from quiescent ones.  No rapidly forming objects
are seen at $z<0.7$ but they become very common at higher redshift.  (b)\
The observed volume production rate of \oii\ (${\cal L}_{\rm
[O\thinspace{II}]}$) as a function of redshift for SSA13 (boxes) and SSA22
(diamonds) compared to the values required to form the present day galaxy
population.  The dashed line corresponds to Gallagher \etal's\cite{gallagher}
calibration of star formation rate versus \oii\ luminosity and the dotted
line to Kennicutt's.\cite{kenn} Some measure of the uncertainties can be
obtained by comparing the two fields.}
\end{figure}

\begin{figure}[h]
\caption{All the 174 magnitude-selected objects in the SSA13 field are
shown in the \bmi\ versus \imk\ color-color plane.  The dashed rectangle
shows the portion of the color-color plane where most of the high-$z$
\oii-luminous galaxies are found.  Many of the unidentified and most of
the unobserved galaxies also lie in this region of color-color space.}
\end{figure}

\begin{figure}[h]
\caption{Deep HST $I$-band images from ref.\ \citen{chain} of all nine
galaxies with $z>1$\ and $L_{\rm [O\thinspace{II}]}\geq\ten{42}$ ergs s$^{-1}$
that lie in the HST fields.  The boxes are $8^{\prime\prime}$\ on a side and
the redshift of each object is shown in the lower right corner.}
\end{figure}

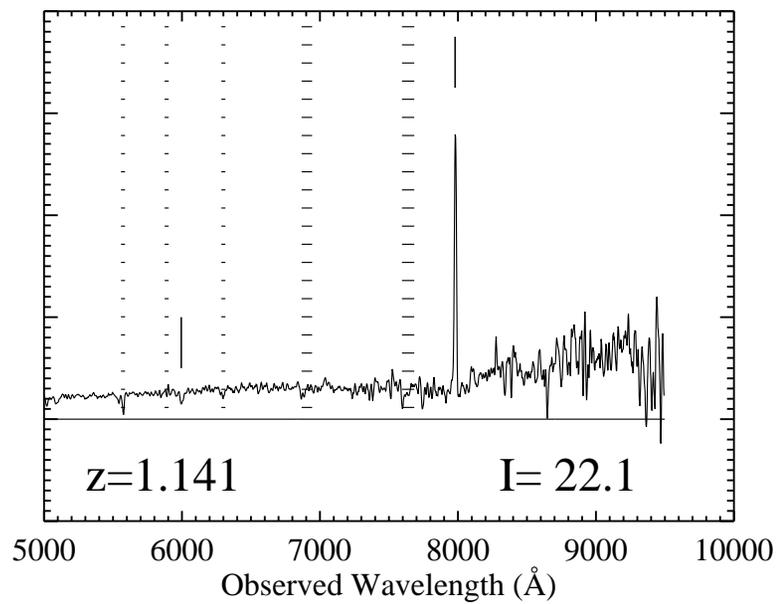
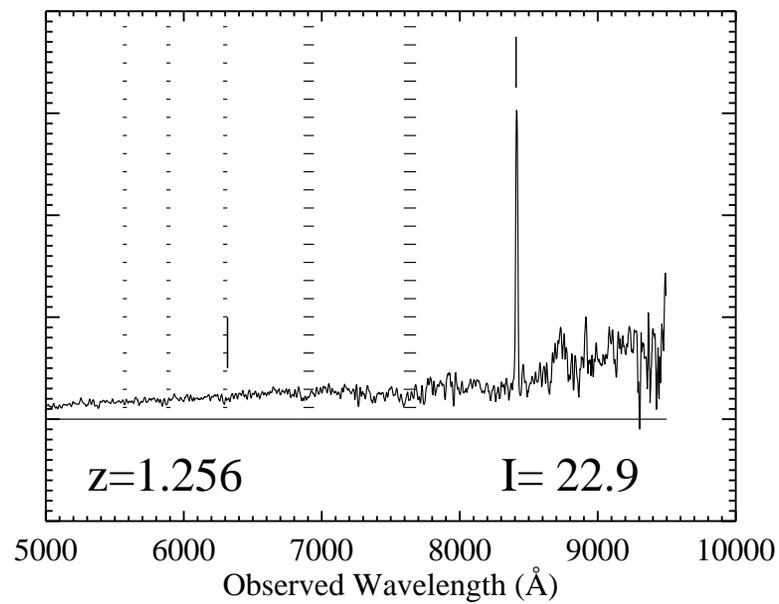
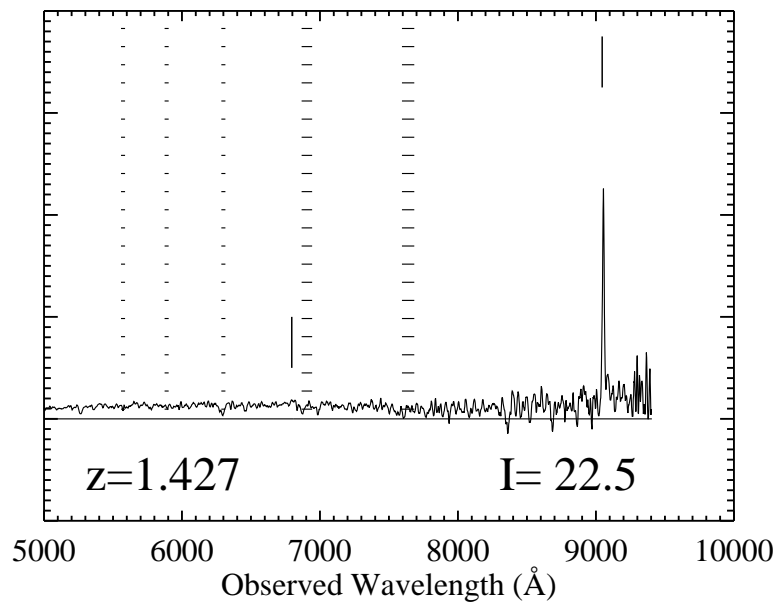
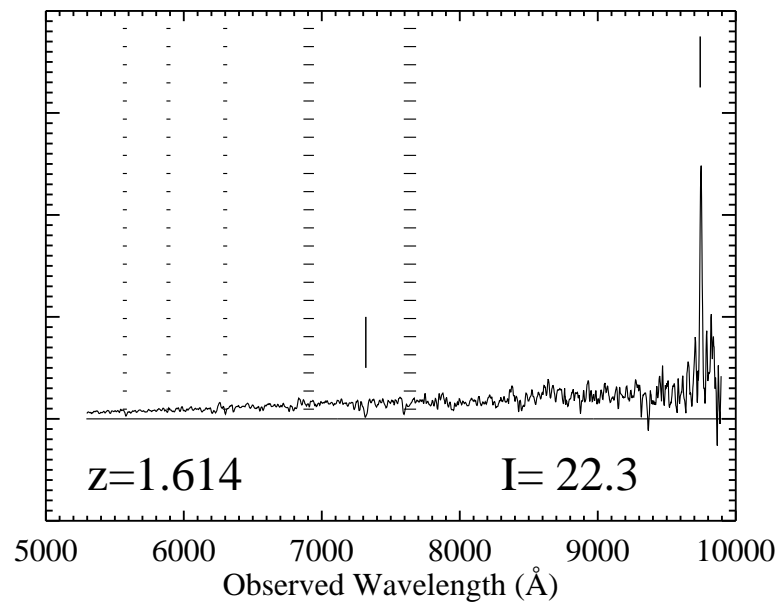
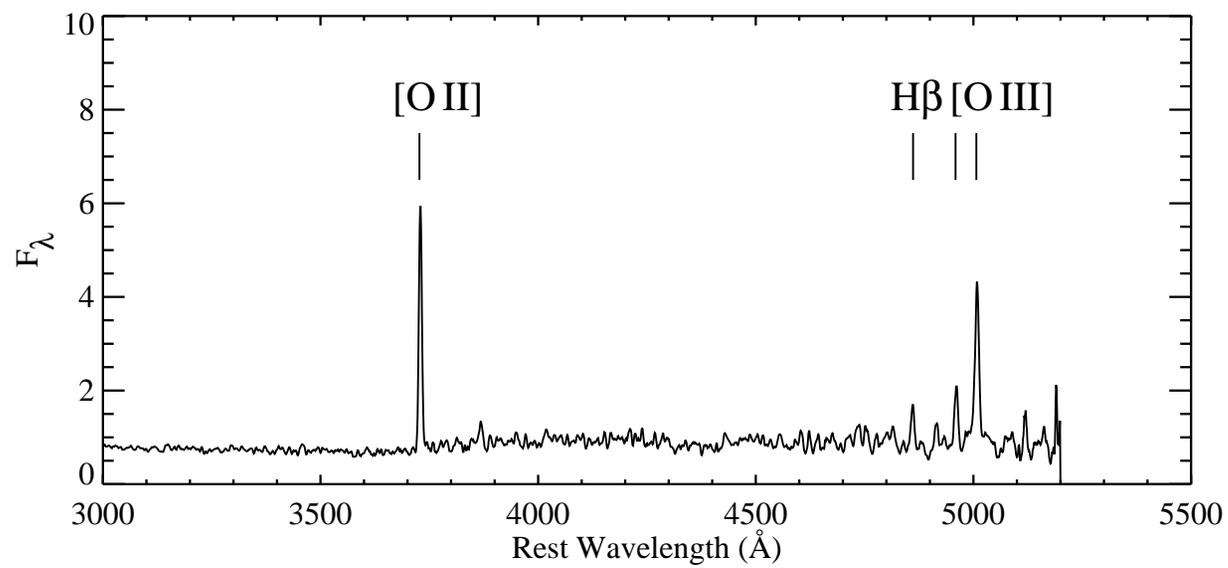
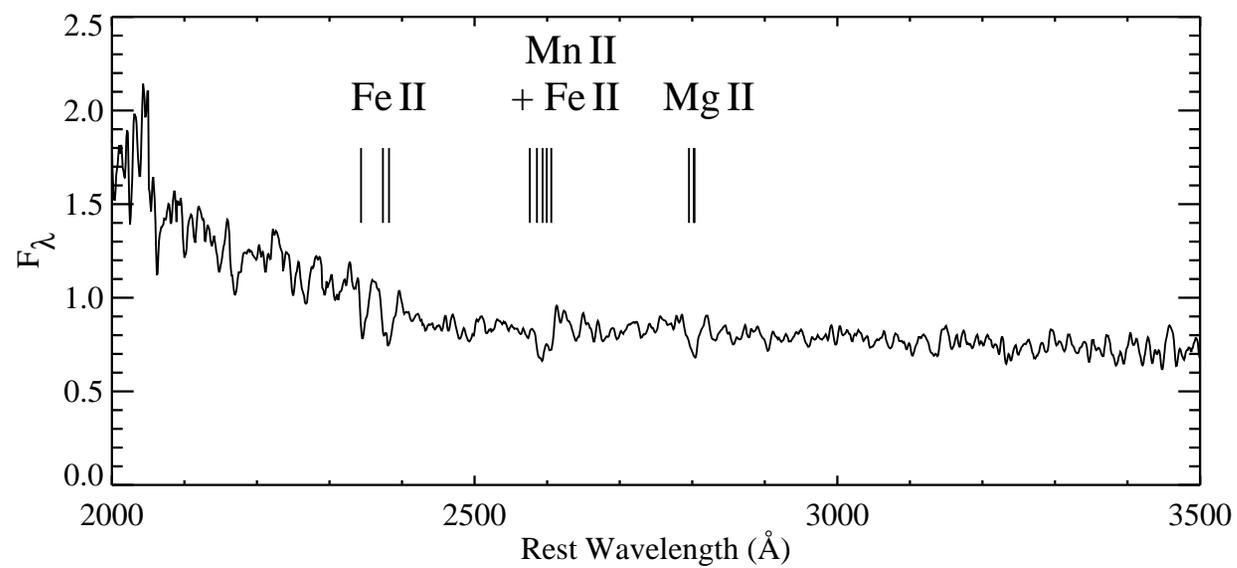
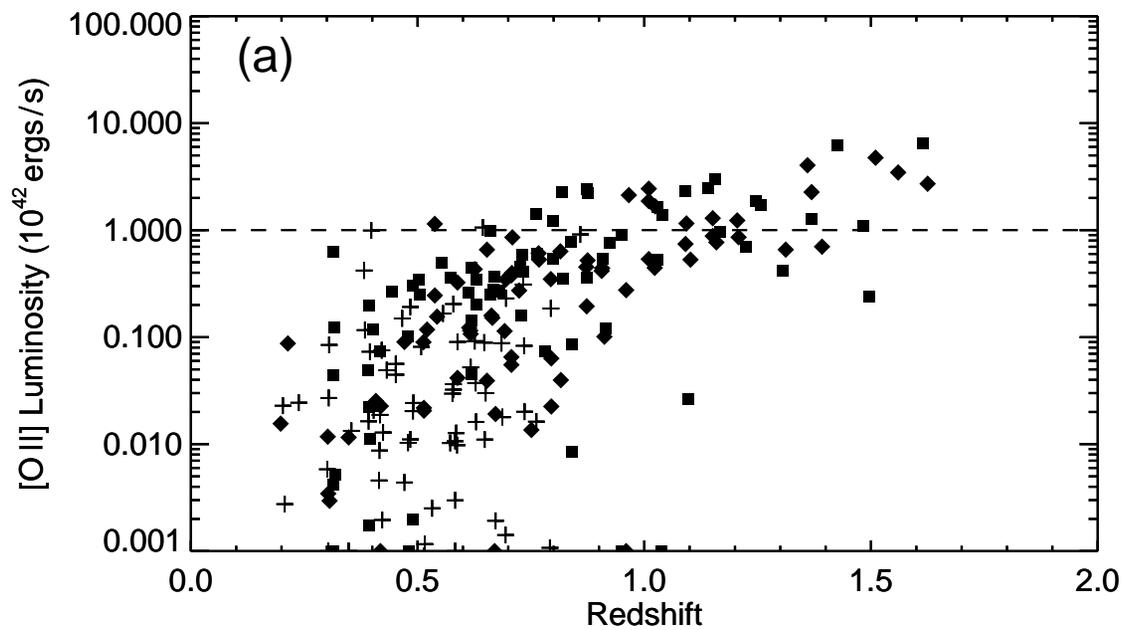
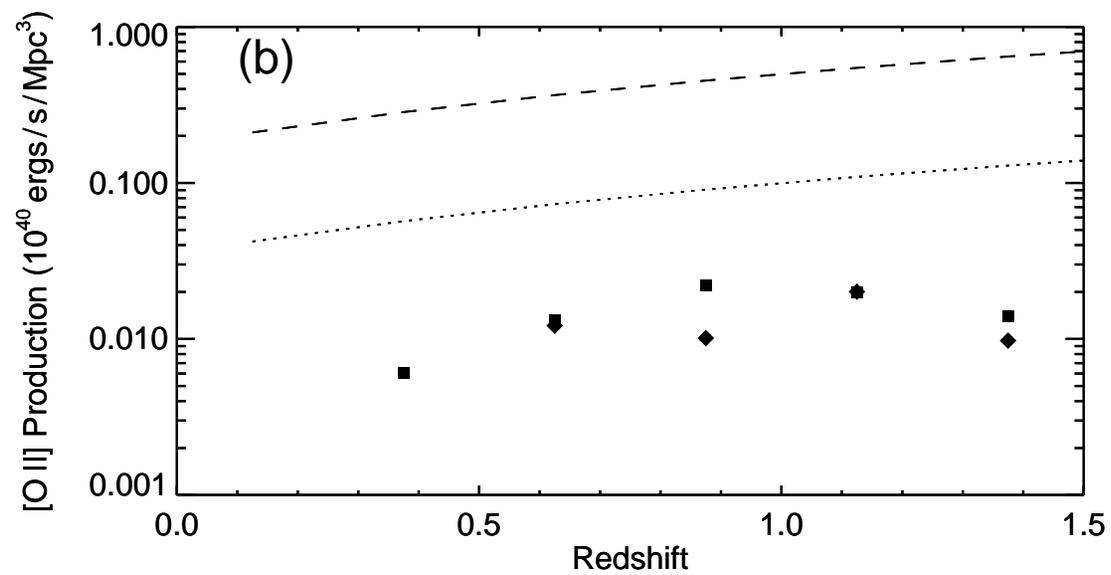
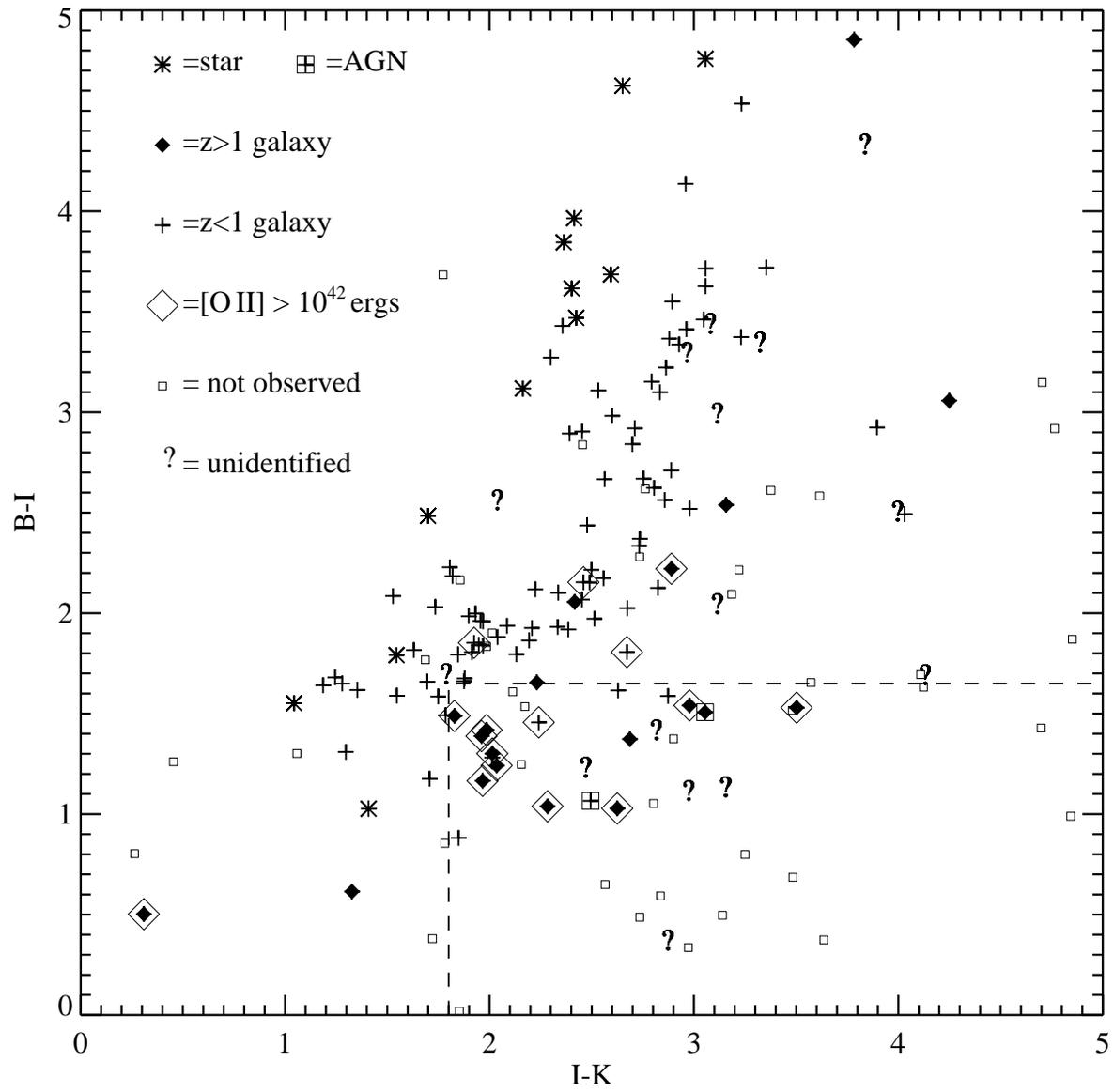
\end{document}